\documentclass[aps,pre,reprint,amsmath,superscriptaddress,showpacs]{revtex4-1}

\usepackage{graphicx}
\usepackage[caption=false]{subfig}
\usepackage[hidelinks]{hyperref}

\begin{document}
\title{Non-parametric estimation of Fisher information from real data}
\date{\today}
\author{Omri \surname{Har Shemesh}}
\email[Electronic mail: ]{O.HarShemesh@uva.nl}

\author{Rick Quax}
\email[Electronic mail: ]{R.Quax@uva.nl}
\affiliation{Computational Science Lab, University of Amsterdam, Science Park 904,
1098XH, Amsterdam, The Netherlands}

\author{Borja Mi\~{n}ano}
\email[Electronic mail: ]{bminyano@mail.iac3.eu}
\affiliation{IAC\textsuperscript{3}–UIB, Mateu Orfila, Cra.\ de Valldemossa %
km 7.5, 07122, Palma, Spain}

\author{Alfons G. Hoekstra}
\email[Electronic mail: ]{A.G.Hoekstra@uva.nl}
\affiliation{Computational Science Lab, University of Amsterdam, Science Park 904,
1098XH, Amsterdam, The Netherlands}
\affiliation{ITMO University, Saint Petersburg, Russia}

\author{Peter M.A. Sloot}
\email[Electronic mail: ]{P.M.A.Sloot@uva.nl}
\affiliation{Computational Science Lab, University of Amsterdam, Science Park 904,
1098XH, Amsterdam, The Netherlands}
\affiliation{ITMO University, Saint Petersburg, Russia}
\affiliation{Complexity Institute, Nanyang Technological University, %
60 Nanyang View, Singapore 639673, Republic of Singapore}

\begin{abstract}

The Fisher Information matrix is a widely used measure for applications ranging
from statistical inference, information geometry, experiment design, to the
study of criticality in biological systems. Yet there is no commonly accepted
non-parametric algorithm to estimate it from real data. In this rapid
communication we show how to accurately estimate the Fisher information in a
non-parametric way. We also develop a numerical procedure to minimize the
errors by choosing the interval of the finite difference scheme necessary to
compute the derivatives in the definition of the Fisher information.  Our
method uses the recently published ``Density Estimation using Field Theory''
algorithm to compute the probability density functions for continuous
densities. We use the Fisher information of the normal distribution to validate
our method and as an example we compute the temperature component of the Fisher
Information Matrix in the two dimensional Ising model and show that it obeys
the expected relation to the heat capacity and therefore peaks at the phase
transition at the correct critical temperature.

\end{abstract}

\maketitle

Consider a probabilistic description of a system as a probability density
function (PDF) $p(x;\theta)$. The observables of the system are collected as
parts of the vector $x$ and the PDF will typically depend on a vector of
continuous parameters $\theta$. It is interesting how the system responds to
parameter changes, e.g.\ at phase transitions or system dynamics near
bifurcations. A measure of the sensitivity of the PDF to the parameters is the
Fisher Information Matrix (FIM)~\cite{Cover2006}. It is a symmetric matrix
labeled by the parameters of the PDF\@. If  a small parameter change causes a
large change in the PDF, the corresponding entry in the FIM will be large. Much
literature discusses the different uses of the FIM, its interpretation as a
Riemmanian metric on the statistical manifold~\cite{Amari} and its relation to
theories of phase transitions~%
\cite{Ruppeiner1979,Ruppeiner1990,Ruppeiner1995,Ruppeiner2012,Ingarden1982,%
Janyszek1989,Janyszek1990a,Brody1995,Brody2009,%
Kumar2012} and complex
systems~\cite{Obata1992,Obata1997,Mayer2006,Frank2009,Prokopenko2011,Wang2011,%
Hidalgo2014}.

Since the FIM is directly computed from the PDF, its accurate estimation
depends on the general density estimation problem~\cite{Silverman1986}.
Density estimation aims to obtain the best estimate $Q_{\mathrm{est}}$ of a
distribution $Q_\mathrm{true}$ given $N$ independently drawn samples. We
distinguish between parametric and non-parametric estimation.  Parametric
estimates constrain $Q_\mathrm{est}$ to depend on a few parameters that are
estimated from the data~\cite{Walter1997}. By the Cram\'er-Rao
inequality~\cite{Cover2006} the inverse of the Fisher information (FI) is a
lower-bound on the variance of the estimated parameters. FI is therefore often
computed in the parametric setting.  In these cases the FI is
computed analytically from the assumed function.  

When we do not assume a specific form for the PDF, we estimate the density
non-parametrically. Thus, the data determines the shape of the distribution.
Areas with higher probability density will contain more data points than areas
with lower probability. The main problem of non-parametric methods is how to
balance the goodness of fit to the data and the smoothness of the estimated
curve~\cite{Silverman1986}. For example, kernel density estimators (KDEs) are a
sum over kernel functions with width $h$, positioned at each data point. i.e.\
$Q_\mathrm{est}(x)=(hN)^{-1}\sum\limits_{i=1}^{N} K[(x-x_i)/h]$ where $x_i$ is
a data point and $K$ is a kernel function.  The bandwidth $h$ controls the
smoothness of the estimate.  In the limit $h\rightarrow 0$ the estimate is a
sum of delta functions at each data point; in the $h\rightarrow\infty$ it is
uniform.  Choosing the correct bandwidth is therefore important.  Taken too
large, the estimate will hide crucial features. Too small a bandwidth causes
spurious peaks in the estimate, especially for long-tailed
distributions~\cite{Silverman1986}. Important to this study, the amount of
smoothing directly affects the value of the FI. This can be
seen from its definition:
\begin{equation} \label{eq:fisher_def} g_{\mu\nu}(\theta) = \langle
  (\partial_\mu\ln p) (\partial_\nu\ln p) \rangle\,, 
\end{equation}
which depends on the derivatives of the PDF\@. Here
$\partial_\mu\equiv\partial/\partial\theta^\mu$, $p$ is a PDF and the average
is with respect to $p$. If, e.g., the estimated PDF $Q_\mathrm{est}$ is
smoother than the true PDF $Q_\mathrm{true}$, the estimate for the FI will be
smaller than the true FI.
 
One elegant approach that derives the smoothness from the data itself was
proposed in~\cite{Bialek1996}. The authors used field theory to formulate the
notion of a smoothness scale as an ultraviolet cutoff, treating the smoothness
length scale $\ell$ as a parameter in a Bayesian inference procedure. They
showed that, in the large $N$ limit, the data selects an appropriate length
scale. Recently, this method was developed into a fast and accurate algorithm
called DEFT (Density Estimation using Field Theory)~\cite{Kinney2014a}. The
algorithm was implemented in one and two dimensions, since it suffers from the
``curse of dimensionality''~\cite{Kinney2014a}. 

Previous work on the nonparametric estimation of FI primarily dealt with PDFs
with location-like parameters, i.e.\ $p(x;\theta)=p(x-\theta)$.  There
Huber~\cite{Huber1974} found a unique interpolation of the cumulative
distribution function that maximizes the FI. Kostal and
Pokora~\cite{Kostal2012} adapted the maximized penalized likelihood method of
Goodd and Gaskins~\cite{Goodd1971} to compute the FI.  Kostal and Pokora
rejected the use of KDE for the direct computation of the FI because no
appropriate bandwidth parameter to control of the $p'/p$ term
in~\eqref{eq:fisher_def} is known~\cite{Kostal2012}.  In this work we estimate
PDFs from samples drawn from the normal distribution at different parameter
values using both DEFT and KDE with a Gaussian kernel, and compare the FI
results to the analytic solution. We obtain accurate results using DEFT, which
are an improvement over using KDE\@. We also use DEFT to compute the FI in the
two dimensional Ising model showing the computation is accurate also at the
phase transition.

To proceed we replace the derivatives in Eq.~\eqref{eq:fisher_def} with
centered finite-difference derivatives:
\begin{subequations}
\label{eq:fisher_finite_diff}
\begin{align}
  g_{\mu\nu}(\theta) &\approx \int \frac{p(x;\theta+\Delta\theta^\mu) - %
  p(x;\theta-\Delta\theta^\mu)}{2\Delta\theta^\mu}\label{eq:finite_deriv}\\%
  \nonumber
&\times\frac{p(x;\theta+\Delta\theta^\nu) -%
p(x;\theta-\Delta\theta^\nu)}{2\Delta\theta^\nu}
\frac{dx}{p(x;\theta)}\\
&\approx \int \frac{\ln p(x;\theta+\Delta\theta^\mu) - %
\ln p(x;\theta-\Delta\theta^\mu)}{2\Delta\theta^\mu}\label{eq:finite_log}\\ 
&\times\frac{\ln p(x;\theta+\Delta\theta^\nu) -
\ln p(x;\theta-\Delta\theta^\nu)}{2\Delta\theta^\nu} p(x;\theta)dx\,.
\nonumber
\end{align}
\end{subequations}
Here $\Delta\theta^\mu$ indicates a change in the value of only one parameter
$\theta^\mu$ keeping all other parameters fixed, i.e., $\theta +
\Delta\theta^\mu \equiv (\theta^1, \ldots, \theta^\mu+\Delta\theta^\mu, \ldots,
\theta^d)$. The error introduced by this replacement is proportional to
$O(\Delta\theta^2/6)$ (for each derivative) as can be verified by Taylor
expansion. Higher order finite-difference schemes can be used but
not, in our experience, a lower order one-sided derivative because 
the estimate does not converge to the true value (data not shown). 

This introduces a new free parameter: the size of the difference
$\Delta\theta^\mu$. The value of $\Delta\theta^\mu$ strongly influences the
accuracy of the computation.  Two sources of error determine the optimal
$\Delta\theta^\mu$: the aforementioned numerical derivative error and the
finite sample size $N$. The first error, which scales like
$O[(\Delta\theta^\mu)^2]$, decreases with decreasing $\Delta\theta$. The second
source, however, decreases with increasing $\Delta\theta^\mu$.  This happens
because an estimate from a finite number of samples is always under-determined.
Any estimate is one curve from a group of curves that are close, but not equal,
to the true density. The larger the number of samples, the smaller the size of
the group. If $\Delta\theta^\mu$ is too small, the groups of the densities in
the numerical derivatives will overlap and the difference
$p(x;\theta+\Delta\theta^\mu) - p(x;\theta-\Delta\theta^\mu)$ will be
ill-defined.  This leads to one of our main results: since $\Delta\theta^\mu$
cannot be too small or too large, there is an optimal value with minimal error
between the two extremes.

\begin{figure}[h]
   \centering
   \subfloat[Well separated densities.]{\label{fig:non_overlapping}%
   \includegraphics[width=0.8\linewidth]{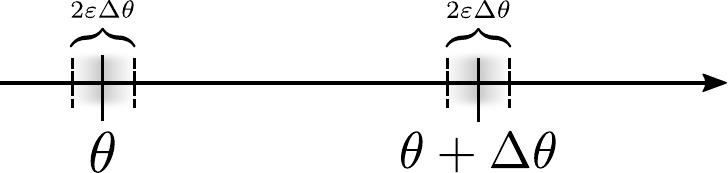}}\quad
   \subfloat[Overlapping densities.]{\includegraphics[width=0.8\linewidth]%
   {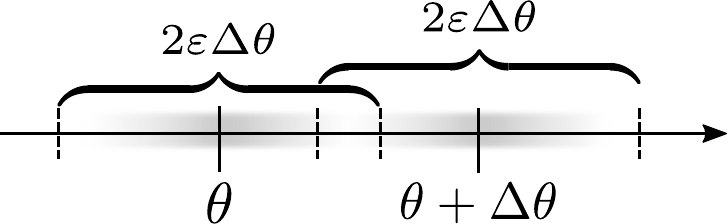}\label{fig:overlapping}}
   \caption[]{Schematic drawing in one dimension with points of estimation
   $\theta$ and $\theta+\Delta\theta$. The gray area is the hypersphere.
   $\varepsilon$ is the radius of the hypersphere in units of
   $\Delta\theta$.}\label{fig:line_drawing}
\end{figure}

To estimate the curve group size and avoid overlaps, we use large deviations
theory. According to Sanov's theorem~\cite{Sanov1957} the appropriate distance
measure is the Kullback-Leibler (KL) divergence:
\begin{equation}
\label{eq:dkl}
\mathcal{D}_{KL}[Q||P] \equiv \int\limits_{x \in \mathcal{X}} Q(x)\ln\frac{Q(x)}{P(x)}dx
\end{equation}
which is defined for two densities $P(x)$ and $Q(x)$ where the support of $P$
and $Q$ overlap. The probability that a set of $N$ samples independently drawn
from $P$ appears to be drawn from $Q$ is proportional to:
\begin{equation}
   \label{eq:sanov1}
   \exp \left(- N \mathcal{D}_{KL}\left[Q||P\right] \right).
\end{equation}
In the limit of infinite sample size this tends to zero. For finite $N$ the set
of distributions whose KL-divergence with $P$ is small enough such that this
probability is finite forms the curve group. This can be interpreted as a
hypersphere in parameter space centered at $\theta$ with an $N$ and $\theta$
dependent radius. To minimize error, the radius at $\theta$ and
$\theta+\Delta\theta^\mu$ should be small compared to $\Delta\theta^\mu$. The
ideal case is drawn schematically in Fig.~\ref{fig:non_overlapping} with well
separated densities and in Fig.~\ref{fig:overlapping} where $\Delta\theta^\mu$
is too small.

To compute hypersphere radius we take $P=p(x;\theta)$ and
$Q=p(x;\theta+\varepsilon\Delta\theta^\mu)$ in Eq.~\eqref{eq:sanov1}. We thus
seek the density $Q$ at the edge of the hypersphere and parametrize it with
$\varepsilon$, the hypersphere radius in units of $\Delta\theta^\mu$.  The
KL-divergence of two neighbouring distributions is approximately:
\begin{equation} \label{eq:kl_approx}
   \mathcal{D}_{KL}[P(\theta)||P(\theta+\varepsilon\Delta\theta^\mu)] \approx
   \frac{\varepsilon^2}{2}g_{\mu\nu}(\theta)\Delta\theta^\mu\Delta\theta^\nu =
   O(\Delta\theta^2) 
\end{equation}
where we use the Einstein summation convention for repeated
indices~\footnote{This well-known result can be derived using a Taylor
expansion and the definition of the FI.}. Inserting
Eq.~\eqref{eq:kl_approx} into Eq.~\eqref{eq:sanov1} we get 
\begin{equation} 
\label{eq:propto} 
\exp \left[-\frac{N\varepsilon^2}{2}
   g_{\mu\nu}(\theta)\Delta\theta^\mu\Delta\theta^\nu\right].  
\end{equation}
We define the boundary of the hypersphere as the point where the probability is
equal to $e^{-1}$. Equating Eq.~\eqref{eq:propto} with $e^{-1}$, we obtain the
radius $\varepsilon$:
\begin{equation}
\label{eq:varepsilon}
\varepsilon^2 = \frac{2}{Ng_{\mu\nu}(\theta)\Delta\theta^\mu\Delta\theta^\nu}\,.
\end{equation}
The radius depends on the number of samples $N$, $\theta$, and
$\Delta\theta^\mu$. At a given $N$ and $\theta$, \emph{increasing}
$\Delta\theta^\mu$ will \emph{decrease} the radius and thus increase accuracy.

As an analytically solvable example, we take
the univariate normal distribution $\mathcal{N}(\mu, \sigma)$.  Its FI is
\begin{equation}
\label{eq:normal_fi}
g_{\mu\mu} = \frac{1}{\sigma^2}; \quad g_{\sigma\sigma} = \frac{2}{\sigma^2};
\quad g_{\mu\sigma} = g_{\sigma\mu} = 0\,.
\end{equation}
We focus on the FI of $\sigma$, which is not a location parameter. Inserting
this to Eq.~\eqref{eq:varepsilon} yields
\begin{equation}
   \label{eq:delta_sigma}
   \Delta\sigma = \sqrt{\frac{2}{\varepsilon^2 Ng_{\sigma\sigma}}} = %
   \frac{\sigma}{\varepsilon\sqrt{N}}\,,
\end{equation}
with $\Delta\sigma\equiv\Delta\theta^\sigma$. This guides the choice of
$\Delta\sigma$ for a given $N$, $\sigma$, and desired radius $\varepsilon$. We
can get the same result using the Cram\'{e}r-Rao inequality. The minimal
variance of an unbiased estimator for $\sigma$ is $1/g_{\sigma\sigma}$. Given
$N$ samples this equals $\sigma^2/2N$.  Demand that the variance of $\sigma$ is
equal to $\frac{1}{2}{(\varepsilon\Delta\sigma)}^2$ (the factor $\frac{1}{2}$
ensures a consistent definition of $\varepsilon$). This variance is equivalent
to a hypersphere radius of $\varepsilon\Delta\sigma$.  We then have
$\sigma^2/2N = {(\varepsilon\Delta\sigma)}^2/2$. Solving for $\Delta\sigma$
yields Eq.~\eqref{eq:delta_sigma}.

For real data the FI is unknown and can be estimated iteratively.  First
compute the FI with $\Delta\theta^\mu$ that ensure a good approximation of the
numerical derivatives. Then use the FI to compute $\varepsilon$. If it is too
large ($\varepsilon\approx 0.1$ seems to be reasonable), increase
$\Delta\theta^\mu$ or $N$.

\begin{figure}[ht]
   \centering
   \includegraphics[width=0.95\linewidth]{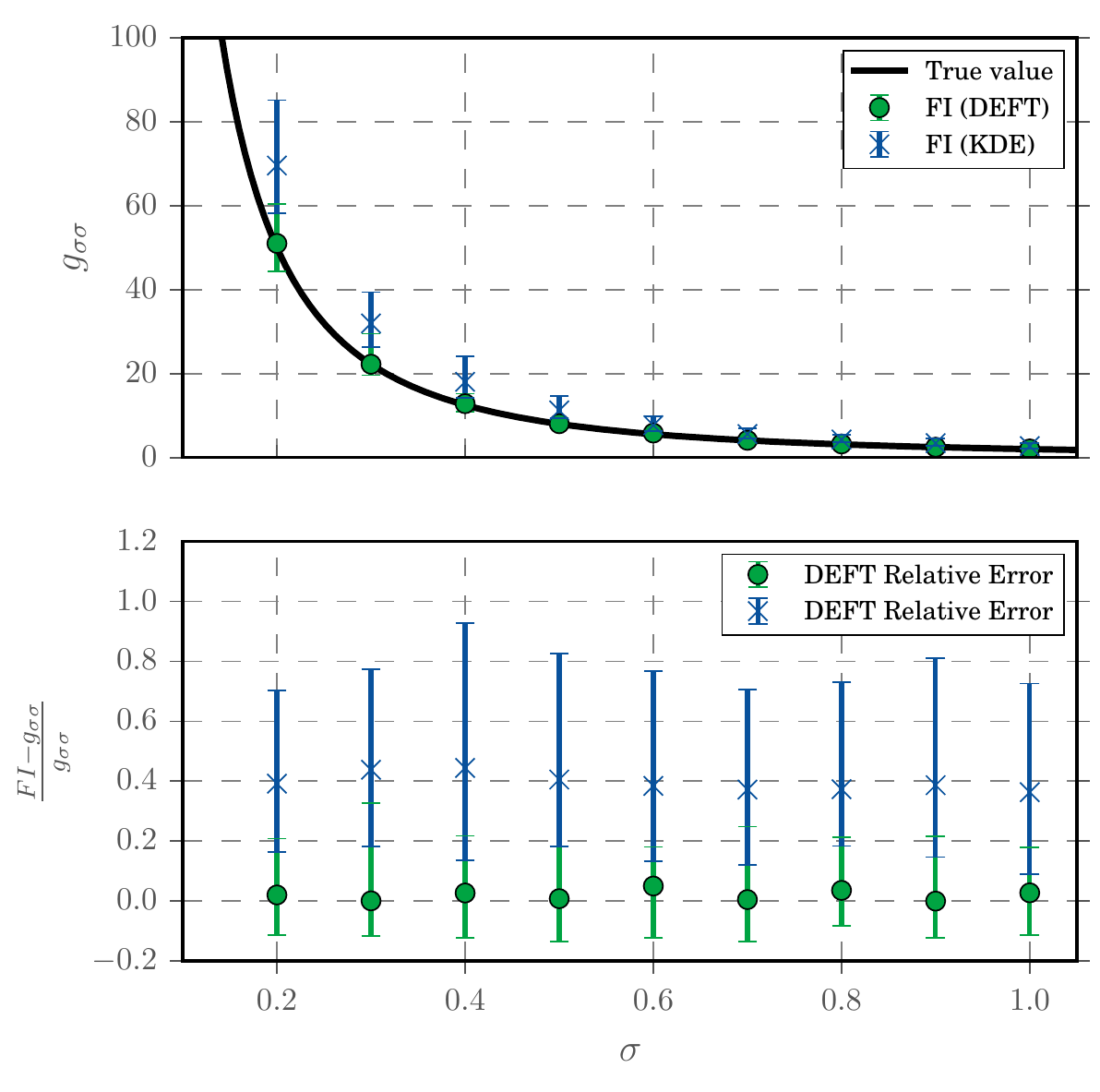}
   \caption{(Color online) A comparison between Gaussian KDE and DEFT for
   density estimation. Top figure shows median FI estimates using both
   methods. Error bars represent $5$ and $95$ percentiles. Bottom figure shows
   the relative errors. The values were computed with $N=10^4$,
   $\varepsilon=0.05$, and $100$ repetitions at each $\sigma$. The same
   samples were used by both methods.}\label{fig:kde_vs_deft}
\end{figure}

We demonstrate our main results by computing $g_{\sigma\sigma}$ from
independently drawn normally distributed samples. We first compare DEFT (with
number of grid points $G=100$, smoothness parameter $\alpha=3$ and a bounding
box twice the interval between the smallest and largest
sample~\cite{Kinney2014a}) and KDE (using Scott's rule for the bandwidth). We
used both with the same samples and computed the FI from
Eq.~\eqref{eq:fisher_finite_diff}.  In the top plot of
Fig.~\ref{fig:kde_vs_deft} the FI estimate is shown.  The black curve is the
analytic value, the green dots and blue $\times$'s are the median estimate
after $100$ repetitions (error bars are $5$ and $95$ percentiles) for DEFT and
KDE respectively. We used $N=10^4$ for each density estimate. We used
$\varepsilon=0.05$ since this yields the best results. 

Both methods seem to follow the analytic curve, however from the relative
errors it is clear that KDE consistently overestimates the FI by about $40\%$
and the distance between $5$ and $95$ percentile is about $100\%$ of the
original value.  DEFT has zero bias and a spread of $30\%-40\%$.  We conclude
that DEFT provides an improvement over KDE both in the estimated value and in
the error margins. In the above computations we used
Eq.~\eqref{eq:finite_deriv} for computation with DEFT and
Eq.~\eqref{eq:finite_log} for KDE, because KDE was extremely unstable when
computed using Eq.~\eqref{eq:finite_deriv} while DEFT performed slightly better
with Eq.~\eqref{eq:finite_deriv}.

\begin{figure}[htb]
   \centering
   \includegraphics[width=\linewidth]{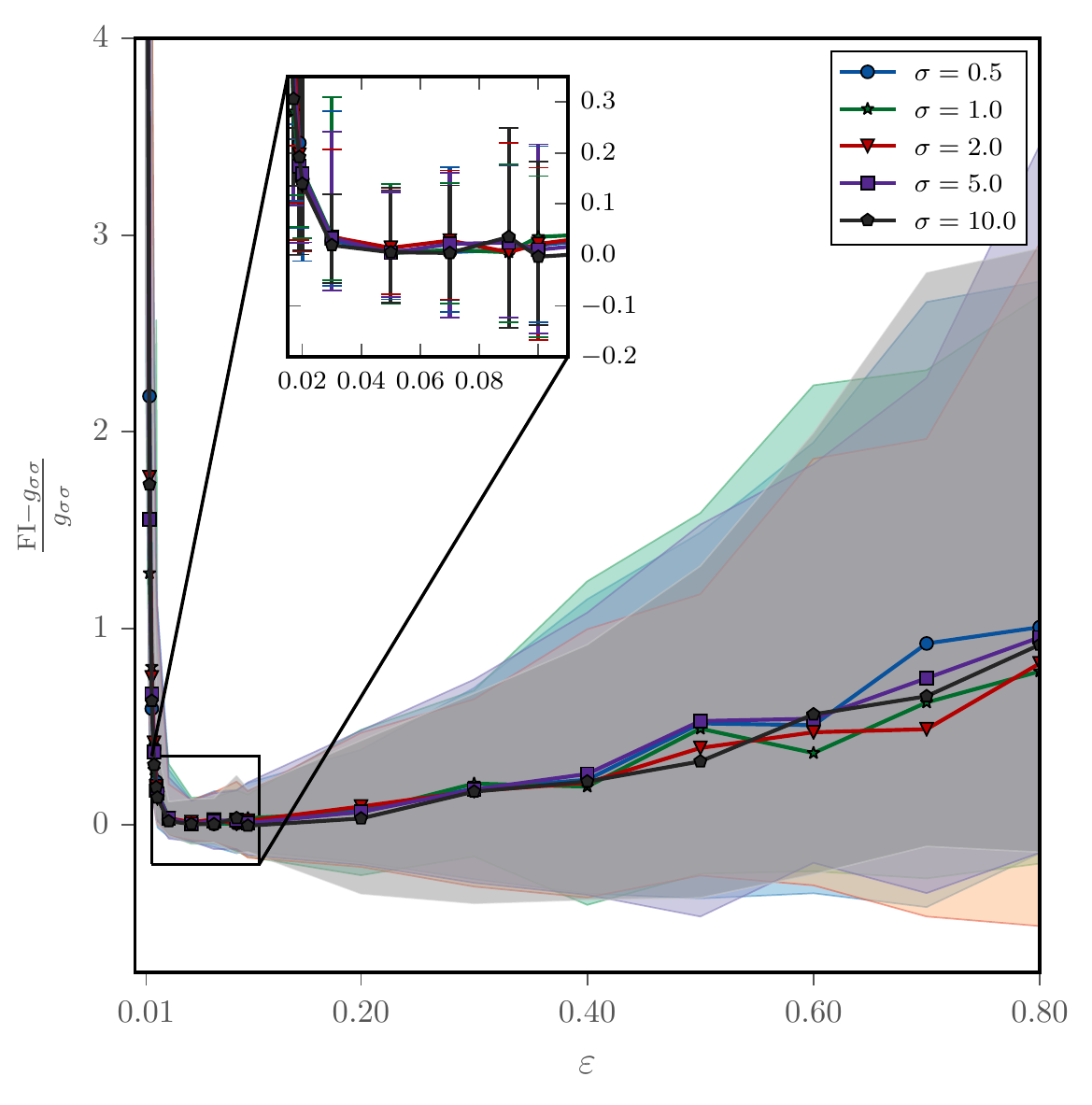}
   \caption{(Color online) The median relative error as a function of
   $\varepsilon$ for different values of $\sigma$. FI stands for the computed
   value and $g_{\sigma\sigma}$ the analytic value. The shaded areas and error
   bars in the inset indicate the $5$ and $95$ percentiles computed over 100
   repetitions of the computation with $N=2\times 10^4$.}
   \label{fig:err_vs_epsilon}
\end{figure}

In the following we use DEFT exclusively for the density estimation.  To see
how the error depends on $\varepsilon$ we vary it at a fixed $N=2\times 10^4$
and plot the relative error.  We computed the FI for $\sigma = 0.5, 1, 2, 5,%
10$. Each computation was repeated $100$ times at different $\varepsilon$ and
the median and $5$ and $95$ percentiles of the relative error
($[g_{\sigma\sigma}-FI]/g_{\sigma\sigma}$, where $FI$ is the estimated FI) were
computed. All curves have the same functional dependence on $\varepsilon$ and,
as we predicted, there is an optimal value for $\Delta\sigma$, at $\varepsilon
\approx 0.05$. Thus the errors depend on $\sigma$ through the combination in
Eq.~\eqref{eq:varepsilon}, as shown in Fig.~\ref{fig:err_vs_epsilon}. All the
curves have a minimum in the range of $\varepsilon \in [0.04, 0.1]$. At small
$\varepsilon$ they grow due to errors in the numerical derivative
($\Delta\sigma$ too large).  At large $\varepsilon$ they grow due to
overlapping densities. The spread (the $90\%$ inter-percentile range) is
minimal at $\varepsilon=0.05$ as well. The shaded regions in the plot represent
the inter-percentile range of the various $\sigma$ curves.

\begin{figure}[htb]
   \centering
   \includegraphics[width=1\linewidth]{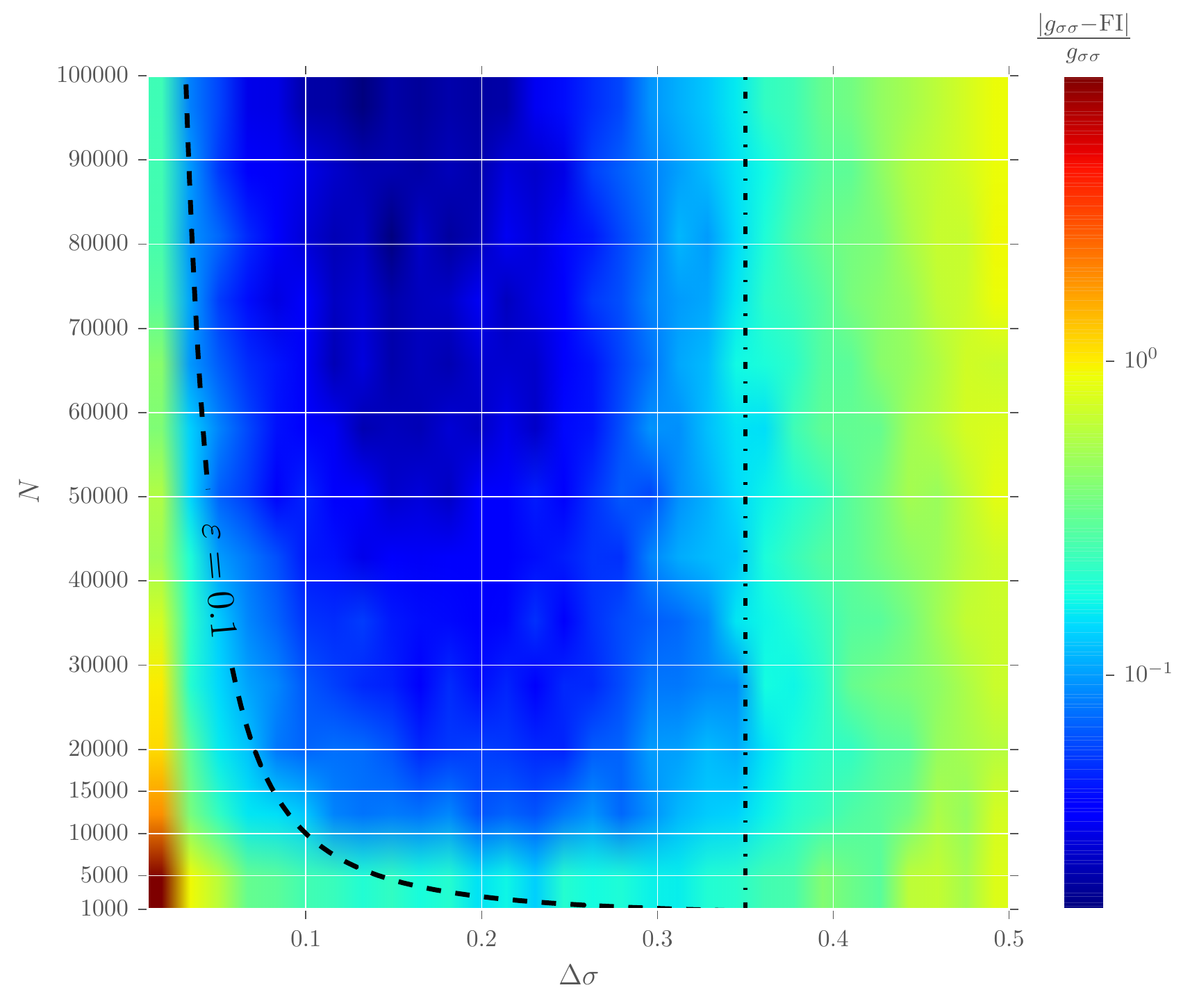}
   \caption{(Color online) Relative error in the computation of the FI for
   $\sigma=1.0$ as a function of both $\Delta\sigma$ and $N$.  Computed using
   DEFT with $100$ repetitions per point. Dashed line represents the
   $\varepsilon=0.1$ line and the dash-dotted line is the $\Delta\sigma=0.35$
   line. Unlike the previous plots, here we compute the absolute value relative
   error to avoid problems with the logarithmic color-bar
   scale.}\label{fig:N_ds_err} 
\end{figure}

To verify the $N$ and $\varepsilon$ dependence of the errors we varied both and
computed $g_{\sigma\sigma}$. The result is presented as a heat map in
Fig.~\ref{fig:N_ds_err}. The color represents the absolute-value relative
estimation error in logarithmic scale. The dashed line indicates the
$\varepsilon=0.1$ line which represents the highest value of $\varepsilon$
where good results are still obtained. The dash-dotted line represents the
$\Delta\sigma=0.35$ line. All computations were done with $\sigma=1.0$ and
$100$ repetitions. The errors due to small $\Delta\sigma$ seem to follow the
$\varepsilon=0.1$ curve, showing again the dependence of this type of error on
$\varepsilon$. Above $\Delta\sigma = 0.35$ we see increasing errors due to the
large value of $\Delta\sigma$. The best area for the estimation is between the
two lines.

One of the applications of the computation of FI from samples
is in detecting phase transitions~\cite{Prokopenko2011}. As a further
validation we took the two dimensional Ising model, which is the prototypical
model of a continuous phase transition. It is a model
of binary spins $s_i$ on a square lattice with nearest-neighbors
interaction. Its Hamiltonian is
\begin{equation}
   \label{eq:ising_hamiltonian}
   \mathcal{H} = -\sum\limits_{\langle i,j\rangle}J_{ij} s_i s_j - %
   h\sum\limits_i s_i\,,
\end{equation}
where $\langle i,j \rangle$ indicates the sum is on nearest neighbors, $s_i =
\pm 1$ is the value of a spin at site $i$, $J_{ij}$ is the interaction energy,
and $h$ is an external applied magnetic field. In more than one dimensions
there is a critical order-disorder phase transition at a finite temperature.
Onsager solved the model exactly in two dimensions in the thermodynamic limit
(infinite number of spins) and at zero applied external
field~\cite{Onsager1944}. The critical temperature in the isotropic case
($J_{ij} \equiv J$) is
\begin{equation}
   \label{eq:critical}
   T_c = \frac{2J}{\ln(1+\sqrt{2})} \approx 2.269J\,.
\end{equation}
For simplicity we set $J \equiv 1$ and Boltzmann's constant $k_B \equiv 1$.

Prokopenko \emph{et.\ al.}~\cite{Prokopenko2011} computed both the $TT$ and
$hh$ components of the FI (computed for the Gibbs distribution
with $\theta^1 = h$ and $\theta^2 = T$) in terms of the susceptibility $\chi_T$
and the specific heat $C_h$ and showed that:
\begin{equation}
\label{eq:g_prop_to}
g_{TT} = \frac{C_h}{T^2}\,; \quad g_{hh} = \frac{\chi_T}{T}\,.
\end{equation}
We therefore expect both to diverge as the system approaches the critical
temperature. In a finite system this means that the FI peaks at
the critical temperature. 

To validate this result we simulate the Ising model and compute the FI\@. We
used the Metropolis-Hastings Monte Carlo algorithm to obtain
samples of the configuration energy with the Gibbs distribution (at zero
external field):
\begin{equation}
\label{eq:gibbs_ising}
p(S;T) = \frac{1}{Z(T)}\exp\left[-\beta\mathcal{H}(S,T,h=0)\right].
\end{equation}
Here $\beta=1/T$ is the inverse temperature, $S={\{s_i\}}_{i=1,\ldots,L^2}$ is
a configuration of the spins on a $L \times L$ square lattice, and $Z$ is the
partition function. We then estimate the $TT$ component of the FI using
Eq.~\eqref{eq:fisher_finite_diff} with densities estimated from the sampled
energies. We also computed the specific heat: 
\begin{equation}
\label{eq:heat_capacity}
   C_h(T) = \frac{1}{L^2T^2}\left( \langle E^2 \rangle - \langle E \rangle^2 \right)
\end{equation}
where $L^2$ is the total number of spins, $E$ is the energy of the
configuration, and the average is performed over different configurations at
the same temperature.

\begin{figure}[htb]
   \centering
   \includegraphics[width=0.9\linewidth]{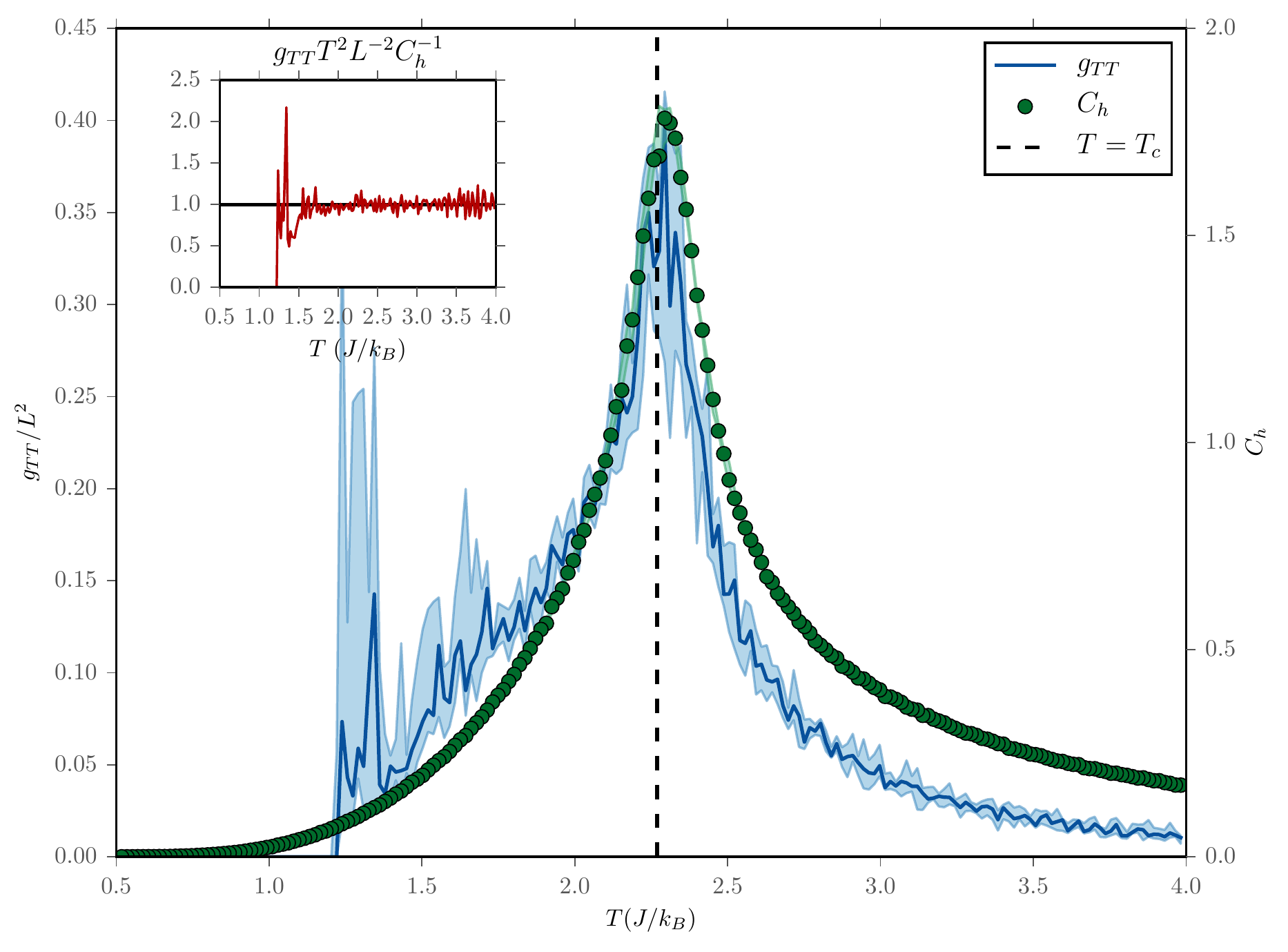}
   \caption{(Color online) Blue continuous curve is the $TT$ component of the
   FIM and the green dots are the heat capacity in the $2D$ Ising model on a
   $25\times 25$ grid.  Shaded blue and green regions indicates the $5$ and
   $95$ percentiles computed from $5$ simulations. Inset shows the ratio of FI
   to heat capacity ($g_{TT}T^2C_v^{-1}L^{-2}$) which according to
   Eq.~\eqref{eq:g_prop_to} is equal to $1$.  }\label{fig:ising}
\end{figure}

We plot the result of both the FI and the specific heat $C_h$ computation in
Fig.~\ref{fig:ising}. The simulation was run on a $25\times 25$ lattice of
spins with periodic boundary conditions in the temperature range $[0.5, 4.0]$
which we divided into $200$ segments, leading to parameter difference of $dT%
\simeq 0.17$. We repeated the simulation $5$ times and compute the median and
$5$ and $95$ percentiles. We used a warm-up period of $5\times 10^{6}$ time
steps and took $N=15,000$ samples of the configuration energy. We used DEFT
(with $G=200$, $\alpha=3$ and a bounding box of $[-4, 1]$) for the density
estimation. Because the FI depends on $T$, $\varepsilon$ was not constant. Its
median was $\varepsilon=0.12 \substack{+0.12 \\ -0.07}$ for the values of
$\varepsilon$ which were not infinite. To verify that Eq.~\eqref{eq:g_prop_to}
holds, we plot the ratio of the two sides of the equation. This is presented in
the inset in Fig.~\ref{fig:ising}.

There are several technical points we would like to mention about the
implementation of the method. First, we performed the same computation with a
smaller grid spacing ($dT=0.007$). This led to a much worse signal-to-noise
ratio because the very close densities caused large peaks to occur, especially
in the low temperature range. Second, it is important to find the most suitable
parameters for DEFT. If the bounding box is too small, or the number of grid
points too small or too large, the estimated density will have multiple peaks
which are not apparent in the data. Thus we recommend plotting the result of
DEFT together with a histogram for several data points to make sure the
convergence is good. Third, in the computation of Eq.~\eqref{eq:finite_deriv}
the term $1/p$ may contribute large values at very small $p$. Equivalently with
Eq.~\eqref{eq:finite_log}, when $p(x|\theta\pm\Delta\theta)$ are small, their
logarithm will again be large. This requires the introduction of a numerical
cutoff. It is common practice to set the contribution of a term where $p(x)=0$
to zero~\cite{Wang2011}. We thus introduced a cutoff such that if any of the
estimates at a particular point is less than the cutoff, the contribution of
this point to the integral will be zero. We investigated the effect of this
cutoff for a range of values between $10^{-20}$ and $10^{-2}$. The value of the
cutoff had very little effect. In the Ising model, the only effect was to
change the size of the low temperature region where the FI is exactly zero (the
lower the cutoff, the smaller the region was). In producing
Fig.~\ref{fig:ising} we used a value of $10^{-10}$. Lastly we would like to
mention that the plots in Fig.~\ref{fig:ising} were obtained by the use of
Eq.~\eqref{eq:finite_log}.

As is clear by the remarks above, care should be taken when using this method
to compute the FI. One should first make sure a good
convergence of DEFT is achieved, by adjusting $G$, $\alpha$ and the bounding
box. Then make sure to select the correct parameter difference $\Delta\theta$,
a decision that can be aided by the estimation of the $\varepsilon$ parameter.
And if necessary, use a cutoff for very low values of the probability
density.  Since we rely on DEFT to perform the density estimation, the
procedure is limited by the limitations of DEFT\@. It is especially important
to note that so far DEFT has been implemented in $1$ and $2$ dimensions. Higher
dimensions suffer from the ``curse of dimensionality'' since they require
exponentially many grid points to evaluate the density.

\begin{acknowledgments}
OHS would like to thank Joan Mass\'{o} and Antoni Arbona from the University of
the Baleric Islands for enlightening discussions. Some of the simulations for
the FI computation in the Ising model were performed using the
Computational Exploratory being developed at the University of the Baleric
Islands. The research leading to these results has received funding from the
European Union Seventh Framework Programme (FP7/2007-2013) under grant
agreement numbers 317534 and 318121.
AgH wishes to acknowledge partial funding by the Russian
Scientific Foundation, under grant \#14-11-00826.  PMAS wishes to acknowledge
partial funding by the Russian Scientific Foundation, under grant
\#14-21-00137.
\end{acknowledgments}


\begin{thebibliography}{29}%
\raggedright
\makeatletter
\providecommand \@ifxundefined [1]{%
 \@ifx{#1\undefined}
}%
\providecommand \@ifnum [1]{%
 \ifnum #1\expandafter \@firstoftwo
 \else \expandafter \@secondoftwo
 \fi
}%
\providecommand \@ifx [1]{%
 \ifx #1\expandafter \@firstoftwo
 \else \expandafter \@secondoftwo
 \fi
}%
\providecommand \natexlab [1]{#1}%
\providecommand \enquote  [1]{``#1''}%
\providecommand \bibnamefont  [1]{#1}%
\providecommand \bibfnamefont [1]{#1}%
\providecommand \citenamefont [1]{#1}%
\providecommand \href@noop [0]{\@secondoftwo}%
\providecommand \href [0]{\begingroup \@sanitize@url \@href}%
\providecommand \@href[1]{\@@startlink{#1}\@@href}%
\providecommand \@@href[1]{\endgroup#1\@@endlink}%
\providecommand \@sanitize@url [0]{\catcode `\\12\catcode `\$12\catcode
  `\&12\catcode `\#12\catcode `\^12\catcode `\_12\catcode `\%12\relax}%
\providecommand \@@startlink[1]{}%
\providecommand \@@endlink[0]{}%
\providecommand \url  [0]{\begingroup\@sanitize@url \@url }%
\providecommand \@url [1]{\endgroup\@href {#1}{\urlprefix }}%
\providecommand \urlprefix  [0]{URL }%
\providecommand \Eprint [0]{\href }%
\providecommand \doibase [0]{http://dx.doi.org/}%
\providecommand \selectlanguage [0]{\@gobble}%
\providecommand \bibinfo  [0]{\@secondoftwo}%
\providecommand \bibfield  [0]{\@secondoftwo}%
\providecommand \translation [1]{[#1]}%
\providecommand \BibitemOpen [0]{}%
\providecommand \bibitemStop [0]{}%
\providecommand \bibitemNoStop [0]{.\EOS\space}%
\providecommand \EOS [0]{\spacefactor3000\relax}%
\providecommand \BibitemShut  [1]{\csname bibitem#1\endcsname}%
\let\auto@bib@innerbib\@empty
\bibitem [{\citenamefont {Cover}\ and\ \citenamefont
  {Thomas}(2006)}]{Cover2006}%
  \BibitemOpen
  \bibfield  {author} {\bibinfo {author} {\bibfnamefont {T.~M.}\ \bibnamefont
  {Cover}}\ and\ \bibinfo {author} {\bibfnamefont {J.~A.}\ \bibnamefont
  {Thomas}},\ }\href
  {http://books.google.com/books?hl=en\&lr=\&id=EuhBluW31hsC\&pgis=1} {\emph
  {\bibinfo {title} {{Elements of Information Theory}}}}\ (\bibinfo
  {publisher} {John Wiley \& Sons},\ \bibinfo {year} {2006})\ p.\ \bibinfo
  {pages} {640}\BibitemShut {NoStop}%
\bibitem [{\citenamefont {Amari}\ and\ \citenamefont {Nagaoka}(2000)}]{Amari}%
  \BibitemOpen
  \bibfield  {author} {\bibinfo {author} {\bibfnamefont {S.-I.}\ \bibnamefont
  {Amari}}\ and\ \bibinfo {author} {\bibfnamefont {H.}~\bibnamefont
  {Nagaoka}},\ }\href@noop {} {\emph {\bibinfo {title} {{Methods of Information
  Geometry; Translations of mathematical monographs, Vol. 191}}}}\ (\bibinfo
  {publisher} {American Mathematical Society},\ \bibinfo {year}
  {2000})\BibitemShut {NoStop}%
\bibitem [{\citenamefont {Ruppeiner}(1979)}]{Ruppeiner1979}%
  \BibitemOpen
  \bibfield  {author} {\bibinfo {author} {\bibfnamefont {G.}~\bibnamefont
  {Ruppeiner}},\ }\href {\doibase 10.1103/PhysRevA.20.1608} {\bibfield
  {journal} {\bibinfo  {journal} {Phys. Rev. A}\ }\textbf {\bibinfo {volume}
  {20}},\ \bibinfo {pages} {1608} (\bibinfo {year} {1979})}\BibitemShut
  {NoStop}%
\bibitem [{\citenamefont {Ruppeiner}\ and\ \citenamefont
  {Davis}(1990)}]{Ruppeiner1990}%
  \BibitemOpen
  \bibfield  {author} {\bibinfo {author} {\bibfnamefont {G.}~\bibnamefont
  {Ruppeiner}}\ and\ \bibinfo {author} {\bibfnamefont {C.}~\bibnamefont
  {Davis}},\ }\href {\doibase 10.1103/PhysRevA.41.2200} {\bibfield  {journal}
  {\bibinfo  {journal} {Phys. Rev. A}\ }\textbf {\bibinfo {volume} {41}},\
  \bibinfo {pages} {2200} (\bibinfo {year} {1990})}\BibitemShut {NoStop}%
\bibitem [{\citenamefont {Ruppeiner}(1995)}]{Ruppeiner1995}%
  \BibitemOpen
  \bibfield  {author} {\bibinfo {author} {\bibfnamefont {G.}~\bibnamefont
  {Ruppeiner}},\ }\href {\doibase 10.1103/RevModPhys.67.605} {\bibfield
  {journal} {\bibinfo  {journal} {Rev. Mod. Phys.}\ }\textbf {\bibinfo {volume}
  {67}},\ \bibinfo {pages} {605} (\bibinfo {year} {1995})}\BibitemShut
  {NoStop}%
\bibitem [{\citenamefont {Ruppeiner}\ \emph {et~al.}(2012)\citenamefont
  {Ruppeiner}, \citenamefont {Sahay}, \citenamefont {Sarkar},\ and\
  \citenamefont {Sengupta}}]{Ruppeiner2012}%
  \BibitemOpen
  \bibfield  {author} {\bibinfo {author} {\bibfnamefont {G.}~\bibnamefont
  {Ruppeiner}}, \bibinfo {author} {\bibfnamefont {A.}~\bibnamefont {Sahay}},
  \bibinfo {author} {\bibfnamefont {T.}~\bibnamefont {Sarkar}}, \ and\ \bibinfo
  {author} {\bibfnamefont {G.}~\bibnamefont {Sengupta}},\ }\href {\doibase
  10.1103/PhysRevE.86.052103} {\bibfield  {journal} {\bibinfo  {journal} {Phys.
  Rev. E}\ }\textbf {\bibinfo {volume} {86}},\ \bibinfo {pages} {052103}
  (\bibinfo {year} {2012})}\BibitemShut {NoStop}%
\bibitem [{\citenamefont {Ingarden}\ \emph {et~al.}(1982)\citenamefont
  {Ingarden}, \citenamefont {Janyszek}, \citenamefont {Kossakowski},\ and\
  \citenamefont {Kawaguchi}}]{Ingarden1982}%
  \BibitemOpen
  \bibfield  {author} {\bibinfo {author} {\bibfnamefont {R.}~\bibnamefont
  {Ingarden}}, \bibinfo {author} {\bibfnamefont {H.}~\bibnamefont {Janyszek}},
  \bibinfo {author} {\bibfnamefont {A.}~\bibnamefont {Kossakowski}}, \ and\
  \bibinfo {author} {\bibfnamefont {T.}~\bibnamefont {Kawaguchi}},\ }\href@noop
  {} {\bibfield  {journal} {\bibinfo  {journal} {Tensor (NS)}\ }\textbf
  {\bibinfo {volume} {37}},\ \bibinfo {pages} {105} (\bibinfo {year}
  {1982})}\BibitemShut {NoStop}%
\bibitem [{\citenamefont {Janyszek}\ and\ \citenamefont
  {Mrugala}(1989)}]{Janyszek1989}%
  \BibitemOpen
  \bibfield  {author} {\bibinfo {author} {\bibfnamefont {H.}~\bibnamefont
  {Janyszek}}\ and\ \bibinfo {author} {\bibfnamefont {R.}~\bibnamefont
  {Mrugala}},\ }\href {\doibase 10.1103/PhysRevA.39.6515} {\bibfield  {journal}
  {\bibinfo  {journal} {Phys. Rev. A}\ }\textbf {\bibinfo {volume} {39}},\
  \bibinfo {pages} {6515} (\bibinfo {year} {1989})}\BibitemShut {NoStop}%
\bibitem [{\citenamefont {Janyszek}(1990)}]{Janyszek1990a}%
  \BibitemOpen
  \bibfield  {author} {\bibinfo {author} {\bibfnamefont {H.}~\bibnamefont
  {Janyszek}},\ }\href {\doibase 10.1088/0305-4470/23/4/017} {\bibfield
  {journal} {\bibinfo  {journal} {J. Phys. A. Math. Gen.}\ }\textbf {\bibinfo
  {volume} {23}},\ \bibinfo {pages} {477} (\bibinfo {year} {1990})}\BibitemShut
  {NoStop}%
\bibitem [{\citenamefont {Brody}\ and\ \citenamefont
  {Rivier}(1995)}]{Brody1995}%
  \BibitemOpen
  \bibfield  {author} {\bibinfo {author} {\bibfnamefont {D.}~\bibnamefont
  {Brody}}\ and\ \bibinfo {author} {\bibfnamefont {N.}~\bibnamefont {Rivier}},\
  }\href {http://pre.aps.org/abstract/PRE/v51/i2/p1006\_1} {\bibfield
  {journal} {\bibinfo  {journal} {Phys. Rev. E}\ }\textbf {\bibinfo {volume}
  {51}},\ \bibinfo {pages} {1006} (\bibinfo {year} {1995})}\BibitemShut
  {NoStop}%
\bibitem [{\citenamefont {Brody}\ and\ \citenamefont {Hook}(2009)}]{Brody2009}%
  \BibitemOpen
  \bibfield  {author} {\bibinfo {author} {\bibfnamefont {D.~C.}\ \bibnamefont
  {Brody}}\ and\ \bibinfo {author} {\bibfnamefont {D.~W.}\ \bibnamefont
  {Hook}},\ }\href {\doibase 10.1088/1751-8113/42/2/023001} {\bibfield
  {journal} {\bibinfo  {journal} {J. Phys. A Math. Theor.}\ }\textbf {\bibinfo
  {volume} {42}},\ \bibinfo {pages} {023001} (\bibinfo {year}
  {2009})}\BibitemShut {NoStop}%
\bibitem [{\citenamefont {Kumar}\ \emph {et~al.}(2012)\citenamefont {Kumar},
  \citenamefont {Mahapatra}, \citenamefont {Phukon},\ and\ \citenamefont
  {Sarkar}}]{Kumar2012}%
  \BibitemOpen
  \bibfield  {author} {\bibinfo {author} {\bibfnamefont {P.}~\bibnamefont
  {Kumar}}, \bibinfo {author} {\bibfnamefont {S.}~\bibnamefont {Mahapatra}},
  \bibinfo {author} {\bibfnamefont {P.}~\bibnamefont {Phukon}}, \ and\ \bibinfo
  {author} {\bibfnamefont {T.}~\bibnamefont {Sarkar}},\ }\href {\doibase
  10.1103/PhysRevE.86.051117} {\bibfield  {journal} {\bibinfo  {journal} {Phys.
  Rev. E}\ }\textbf {\bibinfo {volume} {86}},\ \bibinfo {pages} {051117}
  (\bibinfo {year} {2012})}\BibitemShut {NoStop}%
\bibitem [{\citenamefont {Obata}\ \emph {et~al.}(1992)\citenamefont {Obata},
  \citenamefont {Hara},\ and\ \citenamefont {Endo}}]{Obata1992}%
  \BibitemOpen
  \bibfield  {author} {\bibinfo {author} {\bibfnamefont {T.}~\bibnamefont
  {Obata}}, \bibinfo {author} {\bibfnamefont {H.}~\bibnamefont {Hara}}, \ and\
  \bibinfo {author} {\bibfnamefont {K.}~\bibnamefont {Endo}},\ }\href
  {http://adsabs.harvard.edu/abs/1992PhRvA..45.6997O} {\bibfield  {journal}
  {\bibinfo  {journal} {Phys. Rev. A}\ }\textbf {\bibinfo {volume} {45}},\
  \bibinfo {pages} {6997} (\bibinfo {year} {1992})}\BibitemShut {NoStop}%
\bibitem [{\citenamefont {Obata}\ \emph {et~al.}(1997)\citenamefont {Obata},
  \citenamefont {Oshima},\ and\ \citenamefont {Hara}}]{Obata1997}%
  \BibitemOpen
  \bibfield  {author} {\bibinfo {author} {\bibfnamefont {T.}~\bibnamefont
  {Obata}}, \bibinfo {author} {\bibfnamefont {H.}~\bibnamefont {Oshima}}, \
  and\ \bibinfo {author} {\bibfnamefont {H.}~\bibnamefont {Hara}},\ }\href
  {\doibase 10.1103/PhysRevE.56.213} {\bibfield  {journal} {\bibinfo  {journal}
  {Phys. Rev. E}\ }\textbf {\bibinfo {volume} {56}},\ \bibinfo {pages} {213}
  (\bibinfo {year} {1997})}\BibitemShut {NoStop}%
\bibitem [{\citenamefont {Mayer}\ \emph {et~al.}(2006)\citenamefont {Mayer},
  \citenamefont {Pawlowski},\ and\ \citenamefont {Cabezas}}]{Mayer2006}%
  \BibitemOpen
  \bibfield  {author} {\bibinfo {author} {\bibfnamefont {A.~L.}\ \bibnamefont
  {Mayer}}, \bibinfo {author} {\bibfnamefont {C.~W.}\ \bibnamefont
  {Pawlowski}}, \ and\ \bibinfo {author} {\bibfnamefont {H.}~\bibnamefont
  {Cabezas}},\ }\href {\doibase 10.1016/j.ecolmodel.2005.11.011} {\bibfield
  {journal} {\bibinfo  {journal} {Ecol. Modell.}\ }\textbf {\bibinfo {volume}
  {195}},\ \bibinfo {pages} {72} (\bibinfo {year} {2006})}\BibitemShut
  {NoStop}%
\bibitem [{\citenamefont {Frank}(2009)}]{Frank2009}%
  \BibitemOpen
  \bibfield  {author} {\bibinfo {author} {\bibfnamefont {S.~A.}\ \bibnamefont
  {Frank}},\ }\href {\doibase 10.1111/j.1420-9101.2008.01647.x} {\bibfield
  {journal} {\bibinfo  {journal} {J. Evol. Biol.}\ }\textbf {\bibinfo {volume}
  {22}},\ \bibinfo {pages} {231} (\bibinfo {year} {2009})}\BibitemShut
  {NoStop}%
\bibitem [{\citenamefont {Prokopenko}\ \emph {et~al.}(2011)\citenamefont
  {Prokopenko}, \citenamefont {Lizier}, \citenamefont {Obst},\ and\
  \citenamefont {Wang}}]{Prokopenko2011}%
  \BibitemOpen
  \bibfield  {author} {\bibinfo {author} {\bibfnamefont {M.}~\bibnamefont
  {Prokopenko}}, \bibinfo {author} {\bibfnamefont {J.~T.}\ \bibnamefont
  {Lizier}}, \bibinfo {author} {\bibfnamefont {O.}~\bibnamefont {Obst}}, \ and\
  \bibinfo {author} {\bibfnamefont {X.~R.}\ \bibnamefont {Wang}},\ }\href
  {http://www.ncbi.nlm.nih.gov/pubmed/22181096} {\bibfield  {journal} {\bibinfo
   {journal} {Phys. Rev. E}\ }\textbf {\bibinfo {volume} {84}},\ \bibinfo
  {pages} {041116} (\bibinfo {year} {2011})}\BibitemShut {NoStop}%
\bibitem [{\citenamefont {Wang}\ \emph {et~al.}(2011)\citenamefont {Wang},
  \citenamefont {Lizier},\ and\ \citenamefont {Prokopenko}}]{Wang2011}%
  \BibitemOpen
  \bibfield  {author} {\bibinfo {author} {\bibfnamefont {X.~R.}\ \bibnamefont
  {Wang}}, \bibinfo {author} {\bibfnamefont {J.~T.}\ \bibnamefont {Lizier}}, \
  and\ \bibinfo {author} {\bibfnamefont {M.}~\bibnamefont {Prokopenko}},\
  }\href {\doibase 10.1162/artl\_a\_00041} {\bibfield  {journal} {\bibinfo
  {journal} {Artif. Life}\ }\textbf {\bibinfo {volume} {17}},\ \bibinfo {pages}
  {315} (\bibinfo {year} {2011})}\BibitemShut {NoStop}%
\bibitem [{\citenamefont {Hidalgo}\ \emph {et~al.}(2014)\citenamefont
  {Hidalgo}, \citenamefont {Grilli}, \citenamefont {Suweis}, \citenamefont
  {Mu\~{n}oz}, \citenamefont {Banavar},\ and\ \citenamefont
  {Maritan}}]{Hidalgo2014}%
  \BibitemOpen
  \bibfield  {author} {\bibinfo {author} {\bibfnamefont {J.}~\bibnamefont
  {Hidalgo}}, \bibinfo {author} {\bibfnamefont {J.}~\bibnamefont {Grilli}},
  \bibinfo {author} {\bibfnamefont {S.}~\bibnamefont {Suweis}}, \bibinfo
  {author} {\bibfnamefont {M.~a.}\ \bibnamefont {Mu\~{n}oz}}, \bibinfo {author}
  {\bibfnamefont {J.~R.}\ \bibnamefont {Banavar}}, \ and\ \bibinfo {author}
  {\bibfnamefont {A.}~\bibnamefont {Maritan}},\ }\href {\doibase
  10.1073/pnas.1319166111} {\bibfield  {journal} {\bibinfo  {journal} {Proc.
  Natl. Acad. Sci. U. S. A.}\ }\textbf {\bibinfo {volume} {111}},\ \bibinfo
  {pages} {10095} (\bibinfo {year} {2014})}\BibitemShut {NoStop}%
\bibitem [{\citenamefont {Silverman}(1986)}]{Silverman1986}%
  \BibitemOpen
  \bibfield  {author} {\bibinfo {author} {\bibfnamefont {B.~W.}\ \bibnamefont
  {Silverman}},\ }\href@noop {} {\emph {\bibinfo {title} {{Density estimation
  for statistics and data analysis}}}},\ Vol.~\bibinfo {volume} {26}\ (\bibinfo
   {publisher} {CRC press},\ \bibinfo {year} {1986})\BibitemShut {NoStop}%
\bibitem [{\citenamefont {Walter}\ and\ \citenamefont
  {Pronzato}(1997)}]{Walter1997}%
  \BibitemOpen
  \bibfield  {author} {\bibinfo {author} {\bibfnamefont {E.}~\bibnamefont
  {Walter}}\ and\ \bibinfo {author} {\bibfnamefont {L.}~\bibnamefont
  {Pronzato}},\ }\href@noop {} {\emph {\bibinfo {title} {Commun. Control
  Eng.}}}\ (\bibinfo  {publisher} {Springer Verlag New-York},\ \bibinfo {year}
  {1997})\BibitemShut {NoStop}%
\bibitem [{\citenamefont {Bialek}\ \emph {et~al.}(1996)\citenamefont {Bialek},
  \citenamefont {Callan},\ and\ \citenamefont {Strong}}]{Bialek1996}%
  \BibitemOpen
  \bibfield  {author} {\bibinfo {author} {\bibfnamefont {W.}~\bibnamefont
  {Bialek}}, \bibinfo {author} {\bibfnamefont {C.~G.}~\bibnamefont {Callan}}, \
  and\ \bibinfo {author} {\bibfnamefont {S.~P.}~\bibnamefont {Strong}},\ }\href
  {\doibase 10.1103/PhysRevLett.77.4693} {\bibfield  {journal} {\bibinfo
  {journal} {Phys. Rev. Lett.}\ }\textbf {\bibinfo {volume} {77}},\ \bibinfo
  {pages} {4693} (\bibinfo {year} {1996})},\ \Eprint
  {http://arxiv.org/abs/9607180v1} {arXiv:9607180v1 [arXiv:cond-mat]}
  \BibitemShut {NoStop}%
\bibitem [{\citenamefont {Kinney}(2014)}]{Kinney2014a}%
  \BibitemOpen
  \bibfield  {author} {\bibinfo {author} {\bibfnamefont {J.~B.}\ \bibnamefont
  {Kinney}},\ }\href {\doibase 10.1103/PhysRevE.90.011301} {\bibfield
  {journal} {\bibinfo  {journal} {Phys. Rev. E}\ }\textbf {\bibinfo {volume}
  {90}},\ \bibinfo {pages} {011301} (\bibinfo {year} {2014})}\BibitemShut
  {NoStop}%
\bibitem [{\citenamefont {Huber}(1974)}]{Huber1974}%
  \BibitemOpen
  \bibfield  {author} {\bibinfo {author} {\bibfnamefont {P.~J.}\ \bibnamefont
  {Huber}},\ }\href {\doibase 10.1214/aos/1176342822} {\bibfield  {journal}
  {\bibinfo  {journal} {Ann. Stat.}\ }\textbf {\bibinfo {volume} {2}},\
  \bibinfo {pages} {1029} (\bibinfo {year} {1974})}\BibitemShut {NoStop}%
\bibitem [{\citenamefont {Kostal}\ and\ \citenamefont
  {Pokora}(2012)}]{Kostal2012}%
  \BibitemOpen
  \bibfield  {author} {\bibinfo {author} {\bibfnamefont {L.}~\bibnamefont
  {Kostal}}\ and\ \bibinfo {author} {\bibfnamefont {O.}~\bibnamefont
  {Pokora}},\ }\href {\doibase 10.3390/e14071221} {\bibfield  {journal}
  {\bibinfo  {journal} {Entropy}\ }\textbf {\bibinfo {volume} {14}},\ \bibinfo
  {pages} {1221} (\bibinfo {year} {2012})}\BibitemShut {NoStop}%
\bibitem [{\citenamefont {Goodd}\ and\ \citenamefont
  {Gaskins}(1971)}]{Goodd1971}%
  \BibitemOpen
  \bibfield  {author} {\bibinfo {author} {\bibfnamefont {I.}~\bibnamefont
  {Goodd}}\ and\ \bibinfo {author} {\bibfnamefont {R.}~\bibnamefont
  {Gaskins}},\ }\href {\doibase 10.1093/biomet/58.2.255} {\bibfield  {journal}
  {\bibinfo  {journal} {Biometrika}\ }\textbf {\bibinfo {volume} {58}},\
  \bibinfo {pages} {255} (\bibinfo {year} {1971})}\BibitemShut {NoStop}%
\bibitem [{\citenamefont {Sanov}(1957)}]{Sanov1957}%
  \BibitemOpen
  \bibfield  {author} {\bibinfo {author} {\bibfnamefont {I.~N.}\ \bibnamefont
  {Sanov}},\ }\href {http://repository.lib.ncsu.edu/dr/handle/1840.4/2119
  http://mi.mathnet.ru/msb5043} {\bibfield  {journal} {\bibinfo  {journal}
  {Mat. Sb.}\ }\textbf {\bibinfo {volume} {42(84)}},\ \bibinfo {pages} {11}
  (\bibinfo {year} {1957})}\BibitemShut {NoStop}%
\bibitem [{Note1()}]{Note1}%
  \BibitemOpen
  \bibinfo {note} {This well-known result can be derived using a Taylor
  expansion and the definition of the Fisher information.}\BibitemShut {Stop}%
\bibitem [{\citenamefont {Onsager}(1944)}]{Onsager1944}%
  \BibitemOpen
  \bibfield  {author} {\bibinfo {author} {\bibfnamefont {L.}~\bibnamefont
  {Onsager}},\ }\href {\doibase 10.1103/PhysRev.65.117} {\bibfield  {journal}
  {\bibinfo  {journal} {Phys. Rev.}\ }\textbf {\bibinfo {volume} {65}},\
  \bibinfo {pages} {117} (\bibinfo {year} {1944})}\BibitemShut {NoStop}%
\end{thebibliography}
\end{document}